\documentclass[aps,prl,reprint,showpacs,amsmath,amssymb,superscriptaddress]{revtex4-1}
\usepackage{amsthm,amsmath,amsfonts,amssymb,amsxtra,appendix,bookmark,dsfont,latexsym,epsfig,graphicx}

\makeatletter
\usepackage{hyperref}
\usepackage{delarray,color}
\usepackage{euscript}

\theoremstyle{definition}

\theoremstyle{remark}

\usepackage{color}

\newcommand{\bdm}{\begin{displaymath}}
\newcommand{\edm}{\end{displaymath}}
\newcommand{\bdn}{\begin{eqnarray}}
\newcommand{\edn}{\end{eqnarray}}
\newcommand{\bay}{\begin{array}{c}}
\newcommand{\eay}{\end{array}}
\newcommand{\ben}{\begin{enumerate}}
\newcommand{\een}{\end{enumerate}}
\newcommand{\beq}{\begin{equation}}
\newcommand{\eeq}{\end{equation}}

\newcommand{\R}{\mathbb{R}}

\newcommand{\curl}{\mathrm{curl} \,}

\newcommand{\bx}{\mathbf{x}}
\newcommand{\by}{\mathbf{y}}
\newcommand{\bA}{\mathbf{A}}
\newcommand{\bJ}{\mathbf{J}}

\newcommand{\im}{\mathrm{i}}

\newcommand{\cEaf}{\mathcal{E} ^{\rm af}_\beta}
\newcommand{\Eaf}{E ^{\rm af}_\beta}
\newcommand{\uaf}{u ^{\rm af}_\beta}

\newcommand{\cETF}{\mathcal{E} ^{\rm TF}_\beta}
\newcommand{\ETF}{E ^{\rm TF}_\beta}
\newcommand{\rhoTF}{\rho ^{\rm TF}_\beta}

\newcommand{\ETFr}{E ^{\rm TF}_1}

\DeclareMathOperator{\dotconv}{\raisebox{-3pt}{\textup{\textbf{*}}}}
\newcommand{\bK}{\mathbf{K}}

\date{April, 2019}

\begin{document} 

\title{Vortex patterns in the almost-bosonic anyon gas}

\author{Michele Correggi}
\affiliation{Dipartimento di Matematica ``G. Castelnuovo'', Universit\`{a} degli Studi di Roma ``La Sapienza'', P.le Aldo Moro, 5, 00185, Rome, Italy.}
\email{michele.correggi@gmail.com}

\author{Romain Duboscq}
\affiliation{Institut de Math\'ematiques de Toulouse, Universit\'e Paul Sabatier, 118 route de Narbonne, F-31062 Toulouse Cedex 9, France}
\email{Romain.Duboscq@math.univ-toulouse.fr}

\author{Douglas Lundholm}
\affiliation{KTH Royal Institute of Technology, Department of Mathematics, SE-100 44 Stockholm, Sweden}
\email{dogge@math.kth.se}

\author{Nicolas Rougerie}
\affiliation{Universit\'e Grenoble-Alpes \& CNRS,  LPMMC (UMR 5493), B.P. 166, F-38 042 Grenoble, France}
\email{nicolas.rougerie@lpmmc.cnrs.fr}

\begin{abstract}
We study theoretically and numerically the ground state of a gas of 2D abelian anyons in an external trapping potential. We treat anyon statistics in the magnetic gauge picture, perturbatively around the bosonic end. This leads to a mean-field energy functional, whose ground state  displays vortex lattices similar to those found in rotating Bose-Einstein condensates. A crucial difference is however that the vortex density is proportional to the underlying matter density of the gas. 
\end{abstract}

\pacs{05.30.Pr, 03.75.Hh}

\maketitle

The eventual unambiguous observation of the elusive exotic particles going under the name of \emph{anyons}~\cite{Khare-05,Myrheim-99,Ouvry-07,Wilczek-90} remains one the most exciting prospects of condensed matter physics. Apart from speculative applications~\cite{NayETALSar-08}, much excitement comes from the possibility of observing quasi-particles~\cite{AroSchWil-84,Haldane-83,Halperin-84} who do \emph{not} fall into the ubiquitous dichotomy between bosons and fermions. 

Contrarily to what happens for bosons and fermions, the many-body problem for non-interacting anyons is in general not exactly solvable. Even the most basic putative experiment thus calls for approximating schemes for its interpretation. In particular, in view of several recent proposals to create anyons in cold quantum gases~\cite{ZhaSreGemJai-14,ZhaSreJai-15,CooSim-15,LunRou-16,YakLem-18,UmuMacComCar-18}, it seems desirable to develop numerically amenable models to describe trapped gases of many anyons. In this note we study such a mean-field-type model, obtained in an ``almost bosonic limit''. In previous work we have derived this model from the many-body Hamiltonian (in a somewhat idealized situation, see the details in~\cite{LunRou-15}) and studied a local density approximation thereof~\cite{CorLunRou-16}. Now we aim at a more quantitative analysis, for which we numerically simulate the ground state of the effective functional. Our main finding (see Figures~\ref{fig:dens quad}-\ref{fig:dens quart} below) is that the ground state develops triangular vortex lattices akin to those found in rotating trapped Bose-Einstein condensates~\cite{Aftalion-06,Cooper-08,Fetter-09,CorPinRouYng-11b,CorRou-13} or type II superconductors~\cite{TilTil-90,Tinkham-75,SanSer-07}. The important difference is however that the density of the vortex patterns is directly related to the underlying matter density profile, that one can compute to be of Thomas-Fermi (TF) shape, within local density approximation (LDA).  

We begin by recalling a few facts about the basic many-anyons Hamiltonian before discussing our average-field (or mean-field) approximation. Then we recall some explicit computations one can make using the LDA, that we shall use as benchmarks for the numerical simulations. Finally we discuss briefly our numerical method and present its results. We find an excellent agreement between the numerics and the available exact benchmarks, which gives us confidence in the new findings, namely, the inhomogeneous vortex lattice.

\medskip

\noindent\textbf{Many-body Hamiltonian.} In the so-called magnetic gauge picture, one trades the statistical phase $e^{i\pi \alpha}, \alpha \in[0,2[$ (or better, braiding phase) that the many-anyons wave function picks after particle exchanges for a Aharonov-Bohm-like magnetic flux of intensity $2\pi\alpha$ attached to each particle. Thus the many-body Hamiltonian for a gas of $N$ 2D anyons in an external potential $V$ becomes (in units where $\hbar = c = 1$ and $m = 1/2$, and with $\bx ^{\perp} = (x,y) ^\perp =  (-y,x)$)
\begin{equation}\label{eq:many body}
H_N ^{\alpha} = \sum_{j= 1} ^N \left( -\im \nabla_{\bx_j} + \alpha \sum_{k\neq j} \frac{\left(\bx_j-\bx_k\right) ^{\perp}}{\left|\bx_j-\bx_k\right| ^{2}}\right) ^2 + V (\bx_j)
\end{equation}
acting on bosonic wave functions $\Psi_N \in L^2_{\rm sym} (\R ^{2N})$ symmetric under particle exchanges. This formulation makes readily clear the reason why the problem is not exactly solvable in general: free anyons (that is, with no further interactions beyond the effect of statistics) correspond to interacting bosons. It is formally equivalent to consider the action of $H^{\alpha-1}_N$ on fermionic wave functions, but we do not follow that route for reasons that will become clear below.

\medskip

\noindent\textbf{Average-field approximation.} We are interested in the most basic question for the many-body Hamiltonian~\eqref{eq:many body}: compute the ground state energy and associated ground states. Since the wave functions we act on are bosonic, a basic mean-field approximation suggests itself. For weak interactions~\cite{Lewin-ICMP,Lewin-XEDP-12,LieSeiSolYng-05,Rougerie-spartacus,Rougerie-LMU}, the ground state of a bosonic system is close to a Bose-Einstein condensate (BEC), 
\begin{equation}\label{eq:BEC}
\Psi_N (\bx_1,\ldots,\bx_N) \approx u (\bx_1) \ldots u (\bx_N). 
\end{equation}
This does not depend much on the particular shape of the interactions and in fact also holds~\cite{LunRou-15} for ground states of~\eqref{eq:many body} in the limit $N\to \infty$ with $\alpha \to 0$ being scaled appropriately, 
a limit we refer to as ``almost bosonic''. Note that in reality $\alpha$ is fixed by the type of anyons under consideration. The almost bosonic limit is thus an idealization, akin to the usual mean-field limit of the interacting Bose gas.

Inserting the ansatz~\eqref{eq:BEC} in~\eqref{eq:many body}, the ground state energy and ground states of~\eqref{eq:many body} are found by minimizing the ``average-field'' functional: with $\beta = \alpha N$,    
\begin{align}\label{eq:AF func}
\cEaf [u] &= \int_{\R^2} \left| \left( -\im \nabla + \beta \bA [|u|^2] \right) u \right|^2 + V |u|^2 \nonumber\\
\bA [\rho] (\bx) &= \int_{\R^2} \frac{(\bx-\by) ^{\perp}}{|\bx-\by| ^2} \rho (\by) d\by.
\end{align}
The corresponding magnetic field
$$\curl \beta \bA [|u|^2] = 2\pi \beta |u|^2$$
has a total flux $2\pi \beta$. Hence, regarding the statistics parameter $\alpha$ in~\eqref{eq:many body} as the number of attached flux units per particle, the total magnetic flux $\beta$ (referred to in~\cite{CorLunRou-16,LunRou-15} as the scaled statistics parameter) should be a large number, proportional to the number of particles. We return to this in the next paragraph.

From now on we shall be concerned only with the effective ground state problem
\begin{equation}\label{eq:AF ener}
 \Eaf = \min \left\{ \cEaf [u], \: \int_{\R^2} |u| ^2 = 1 \right\} = \cEaf [\uaf], 
\end{equation}
which means that we expect our results to be quantitatively valid for anyons with small statistics parameter $\alpha = \beta / N $. However, we expect the qualitative findings to be general. 

Note the kinship of this formalism with that of Chern-Simons theory, see~\cite{Enger-thesis,IenLec-92,Rajaratnam-thesis} and references therein. In~\cite{EdmValOhb-15,ButValOhb-16,ButValOhb-16b} it is also proposed that density-dependent synthetic gauge fields can be created by submitting a Bose gas to both rotation and laser-matter coupling. This leads to a mean-field model with a density-dependent $\bA$ as above, but with a different density dependence $\bA [\rho] \propto \rho \nabla \phi$ where $\phi$ is the laser's phase.

 \begin{figure}
\begin{center}
\begin{minipage}{4.25cm}
\includegraphics[width=4.25cm]{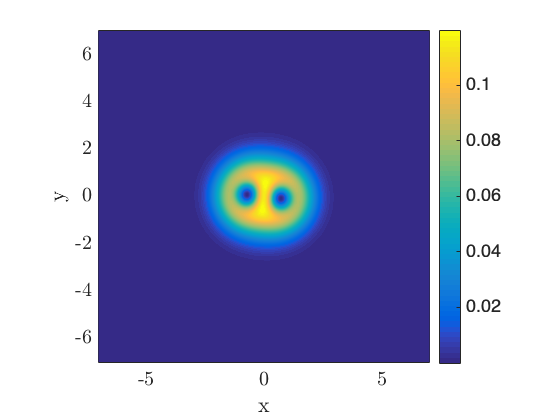}\\
$\beta = 5$
\end{minipage}
\begin{minipage}{4.25cm}
\includegraphics[width=4.25cm]{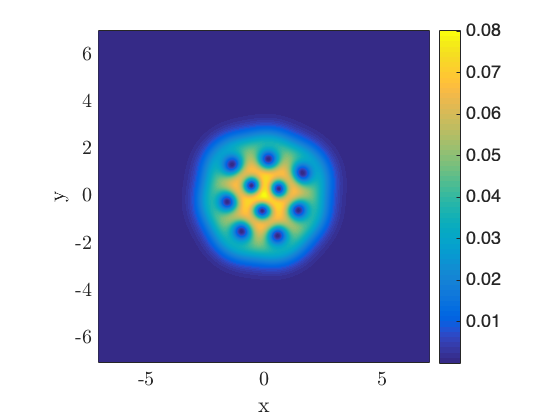}\\
$\beta = 15$
\end{minipage}
\begin{minipage}{4.25cm}
\includegraphics[width=4.25cm]{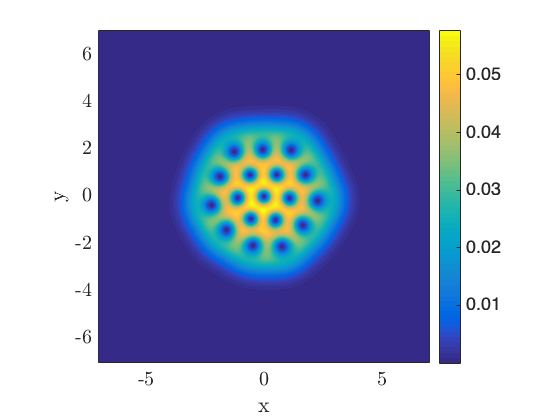}\\
$\beta = 25$
\end{minipage}
\begin{minipage}{4.25cm}
\includegraphics[width=4.25cm]{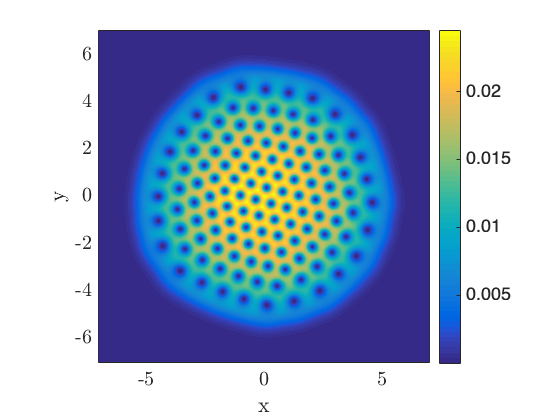}\\
$\beta = 140$
\end{minipage}
%
%
%
\caption{Mean-field approximation $|\uaf| ^2$ for the density of an anyon gas with total magnetic flux $\beta$  in a quadratic trap $V(\bx) = |\bx| ^2$, at $\beta= 5,15,25,140$.}
\label{fig:dens quad}
\end{center}
 \end{figure}

 \begin{figure}
\begin{center}
\begin{minipage}{4.25cm}
\includegraphics[width=4.25cm]{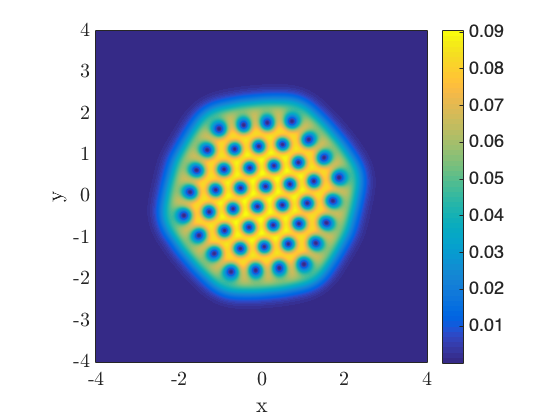}\\
$\beta = 55$
\end{minipage}
\begin{minipage}{4.25cm}
\includegraphics[width=4.25cm]{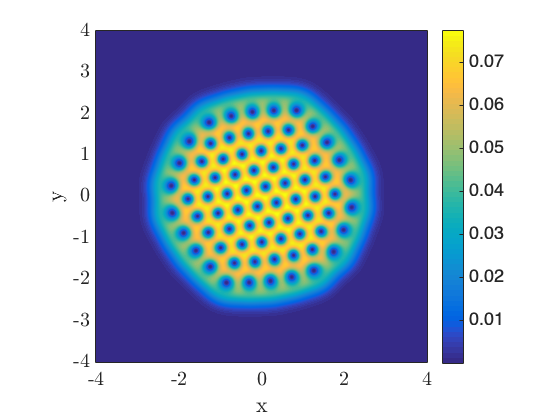}\\
$\beta = 90$
\end{minipage}
\begin{minipage}{4.25cm}
\includegraphics[width=4.25cm]{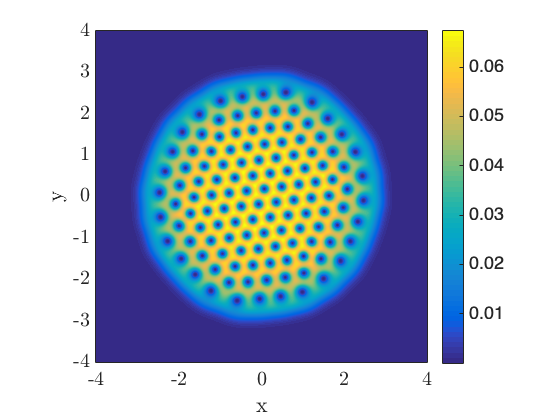}\\
$\beta = 140$
\end{minipage}
\begin{minipage}{4.25cm}
\includegraphics[width=4.25cm]{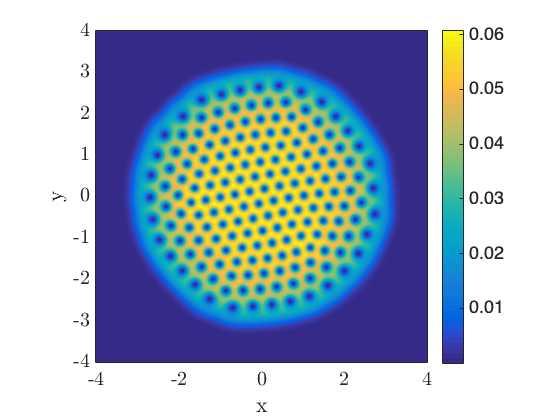}\\
$\beta = 195$
\end{minipage}
\caption{Mean-field approximation $|\uaf| ^2$ for the density of an anyon gas with total magnetic flux $\beta$  in a quartic trap $V(\bx) = |\bx| ^4$, at $\beta= 55,90,140,195$.}
\label{fig:dens quart}
 \end{center}
 \end{figure}

\medskip

\noindent\textbf{Local density approximation.} In view of its relation with the original statistics parameter, $\beta$ should actually be a large parameter in the $N\to \infty$ limit. This suggests looking at the limit $\beta \to \infty$ of Problem~\eqref{eq:AF ener}. Modulo a possible issue~\cite{CorLunRou-16,LunRou-15,Lundholm-16} about commuting the $N\to \infty$ and $\beta \to \infty$ limits, this means the statistics parameter moves away from the bosonic end (``less bosonic'' anyons).  

In this situation, the self-consistent magnetic interaction in~\eqref{eq:AF func} will make the gas expand, so that the first approximation that comes to mind is a LDA. We expect (and, in fact, proved in~\cite{CorLunRou-16}) that in such a limit 
\begin{align}\label{eq:ener large beta}
\Eaf &\underset{\beta \to \infty}{\sim} \ETF\nonumber\\
|\uaf| ^2 &\underset{\beta \to \infty}{\sim} \rhoTF
\end{align}
where $\ETF$ is the local density approximation of the ground state energy: 
\begin{align}\label{eq:TF ener}
\ETF &= \min \left\{ \cETF [\rho], \: \rho \geq 0, \int_{\R^2} \rho = 1 \right\} = \cETF [\rhoTF] \nonumber\\
\cETF [\rho] &= \int_{\R^2} \left( e (\beta,\rho (\bx)) + V(\bx) \rho(\bx) \right) d\bx. 
\end{align}
Here $\rho(x)$ plays the role of the local matter density $|u(x)|^2$ and $e(\beta,\rho)$ is the thermodynamic limit~\cite{CorLunRou-16} of the ground state energy of the homogeneous analogue of~\eqref{eq:AF func}, at average density $\rho$ and total magnetic flux $\beta$. It follows from simple scaling considerations that 
$$ e(\beta, \rho) = \beta \rho ^2 e(1,1).$$
Thus 
$$ \cETF [\rho] = \int_{\R^2} \left( \beta e (1,1) \rho (\bx) ^2  + V(\bx) \rho(\bx) \right) d\bx.$$
The above is similar to the TF functional often used to describe BECs in first approximation (which in turn is, in a rather formal way, similar to the true TF functional for electrons). Note that $\ETF$ depends on a single unknown parameter $e(1,1)$, the thermodynamic ground state energy per unit area at density $\rho = 1$ and magnetic flux $\beta=1$.

\medskip

\noindent\textbf{Energy and density profile.} The TF ground state problem is exactly solvable, modulo the unknown parameter $e(1,1)$. One finds 
\begin{equation}\label{eq:TF profile}
 \rhoTF (\bx) = \frac{1}{2\beta e (1,1)}\left( \lambda_{\beta} ^{\rm TF} - V (\bx)\right)_+  
\end{equation}
where the chemical potential $\lambda_{\beta} ^{\rm TF}$ is set by the normalization $\int_{\R^2} \rhoTF = 1$ and $( \, . \, )_+$ stands for the positive part. More explicitly, we shall in the sequel restrict to radial power-law traps 
\begin{equation}\label{eq:trap}
V (\bx) = |\bx| ^s, \quad s >0. 
\end{equation}
The length scale of the cloud is then $\sim \beta ^{1/(s+2)}$ in the limit $\beta \to \infty$ and, by scaling, 
\begin{equation}\label{eq:TF scaling}
\ETF = \beta ^{s/(s+2)} \ETFr 
\end{equation}
in terms of the energy at $\beta = 1$
$$ \ETFr = \frac{1}{2}\frac{s}{s+1} \left(\frac{s+2}{s}\right)^{2\frac{s+1}{s+2}} \left(\frac{2e(1,1)}{\pi}\right)^{\frac{s}{s+2}},
$$
which can be combined with~\eqref{eq:TF scaling} to recover $e(1,1)$, given $\ETF$:
\begin{equation}\label{eq:e11 estimate general}
	e(1,1) = \frac{\pi}{2} \left(2\frac{s+1}{s}\right)^{\frac{s+2}{s}} \left(\frac{s}{s+2}\right)^{2\frac{s+1}{s}} \frac{(\ETF)^{\frac{s+2}{s}}}{\beta}.
\end{equation}

\medskip

\noindent\textbf{Vortex density.} Concerning the phase of the wave function $\uaf$ we expect the appearance of quantized vortices with average vorticity density given by 
\begin{equation}\label{eq:vorticity}
\boxed{\mu_v = - 2\pi \beta |\uaf|^2 \approx - 2\pi \beta \rhoTF}. 
\end{equation}
Actually, the very fact that a result such as~\eqref{eq:ener large beta} can hold is an indication that the system nucleates quantized vortices. Indeed, for the LDA to be acceptable, it must be that the long range forces encoded in the magnetic vector potential $\bA [|u|^2]$ of~\eqref{eq:AF func} are screened. The physics is thus that a pattern of phase circulation is developed by the ground state on a ``microscopic'' length scale. This phase circulation cancels the long-range component of $\bA[|u|^2]$ and allows for the TF profile to emerge on the macroscopic length scale $\beta ^{1/(s+2)}$. We refer to~\cite{CorLunRou-16} for details, in particular for the discussion of a trial state giving the correct energy by developing phase circulations on length scales $O (\beta^{-s/(2(s+2))})$.   

It is intuitively clear that the phase circulation responsible for the validity of the LDA must come from quantized vortices in the gas. Indeed, if one writes the magnetic kinetic energy of~\eqref{eq:AF func} in the manner 
$$ 
\int_{\R^2} |\nabla \sqrt{\rho}| ^2 + \rho \left| \nabla \varphi + \beta \bA [\rho] \right|^2
$$
by separating density $\rho$ and phase $\varphi$, minimization of the second term would suggest 
$$ \nabla \varphi \approx - \beta \bA [\rho]$$
and thus, taking the curl,
$$ \curl \nabla \varphi \approx - 2\pi \beta \rho$$
which is possible only if $\varphi$ has singularities with non-trivial circulation, i.e. vortices, distributed according to~\eqref{eq:vorticity}. Just as in a rotating Bose gas~\cite{Aftalion-06,Cooper-08,Fetter-09}, we expect that the vortices are singly quantized, for the self-energy of a vortex of degree $d$ should be proportional to $d^2$.

Another argument is as follows. Consider the energy $E^{\rm af} (\beta, M)$ of a homogeneous anyon gas of total mass $M$ in a fixed container $\Omega$. Minimization of the functional~\eqref{eq:AF func} leads to the variational equation
\begin{multline} \label{eq:AF-equation} 
		\left( -i\nabla + \beta\bA[|u|^2] \right)^2 u \\
		-2\beta \frac{\bx^\perp}{|\bx|^2} \dotconv \left( \beta\bA[|u|^2]|u|^2 + \bJ[u] \right)
		u = \lambda u,
\end{multline}
with the current
\begin{equation}\label{eq:J}
\bJ[u] := \frac{i}{2}\left( u\nabla\bar{u} - \bar{u}\nabla u \right),
\end{equation}
and the chemical potential (Lagrange multiplier) 
$$\lambda = \frac{\partial E^{\rm af} (\beta,M)}{\partial M}.$$
In the case of a fixed container,~\eqref{eq:ener large beta}-\eqref{eq:TF scaling} yield $ E^{\rm af} (\beta,M) \propto \beta M^2$, thus we expect that $\lambda \propto \beta M$. 
But, multiplying~\eqref{eq:AF-equation} by $\bar{u}$ and integrating yields (boundary terms vanish because of magnetic Neumann boundary conditions)
\begin{multline} \label{eq:homog start}
	\int_\Omega \left| \left( -i\nabla +  \beta \bA [|u|^2] \right) u \right|^2
	- \lambda \int_\Omega |u|^2
	\\ = 2\beta^2 \int_\Omega |u|^2 \left(\frac{\bx^{\perp}}{|\bx| ^2} \dotconv \bK[u] \right),
\end{multline}
with 
$$
	\bK[u] := \bA[|u|^2]|u|^2 + \frac{1}{\beta}\bJ[u].
$$
The left hand side of~\eqref{eq:homog start} is $O(\beta)$ so we should expect, for large $\beta$
\begin{multline*}
	\int_\Omega |u|^2 \left(\frac{\bx^{\perp}}{|\bx| ^2} \dotconv \bK[u] \right) 
	= \int_\Omega |u|^2 \left(\log |\bx| * \mbox{curl } \bK[u]\right)\\
	= O(\beta ^{-1}).
\end{multline*}
But, on a length scale where the density $|u| ^2$ is homogeneous enough we may approximate 
$$
	\curl \bK[u]  \approx 2\pi |u|^4 + \frac{|u| ^2}{\beta} \curl \nabla \varphi,
$$
so that the above again suggests that~\eqref{eq:vorticity} must hold on sufficiently large scales. We vindicate this expectation in the sequel, by numerical simulations of the full ground state problem.

\medskip

\noindent\textbf{Numerical method.} To minimize~\eqref{eq:AF func} we approximate the evolution in imaginary time  of an initial trial state. The actual implementation is by a pre-conditioned conjugate gradient method \cite{AntLeviTang-17,DanPro-17} using the differential of the energy functional (see~\eqref{eq:AF-equation}), with a projection step ensuring the preservation of the mass. The most costly task is to compute the non-linear non-local terms in~\eqref{eq:AF-equation}. Since they have the form of convolutions, we find it more convenient to work in Fourier space. Our numerical window is thus equipped with periodic boundary conditions and we use a Fast Fourier Transform with respect to a Cartesian grid, so that the non-linearities are simply dealt with by products in the Fourier domain. Note that the kernel defining $\bA [\rho]$ is singular and long range, which forces us to use a cut-off in the space variable and a rather large computational domain in the Fourier space to evaluate it.
 
\medskip

\noindent\textbf{Numerical results: energy and density.} We first plot the density in color levels for two model choices of the trapping potential in Figures~\ref{fig:dens quad} and \ref{fig:dens quart}. The main virtue of these plots is to make the vortex lattice apparent. 

\begin{figure}
\begin{center}
 \includegraphics[width=5.5cm]{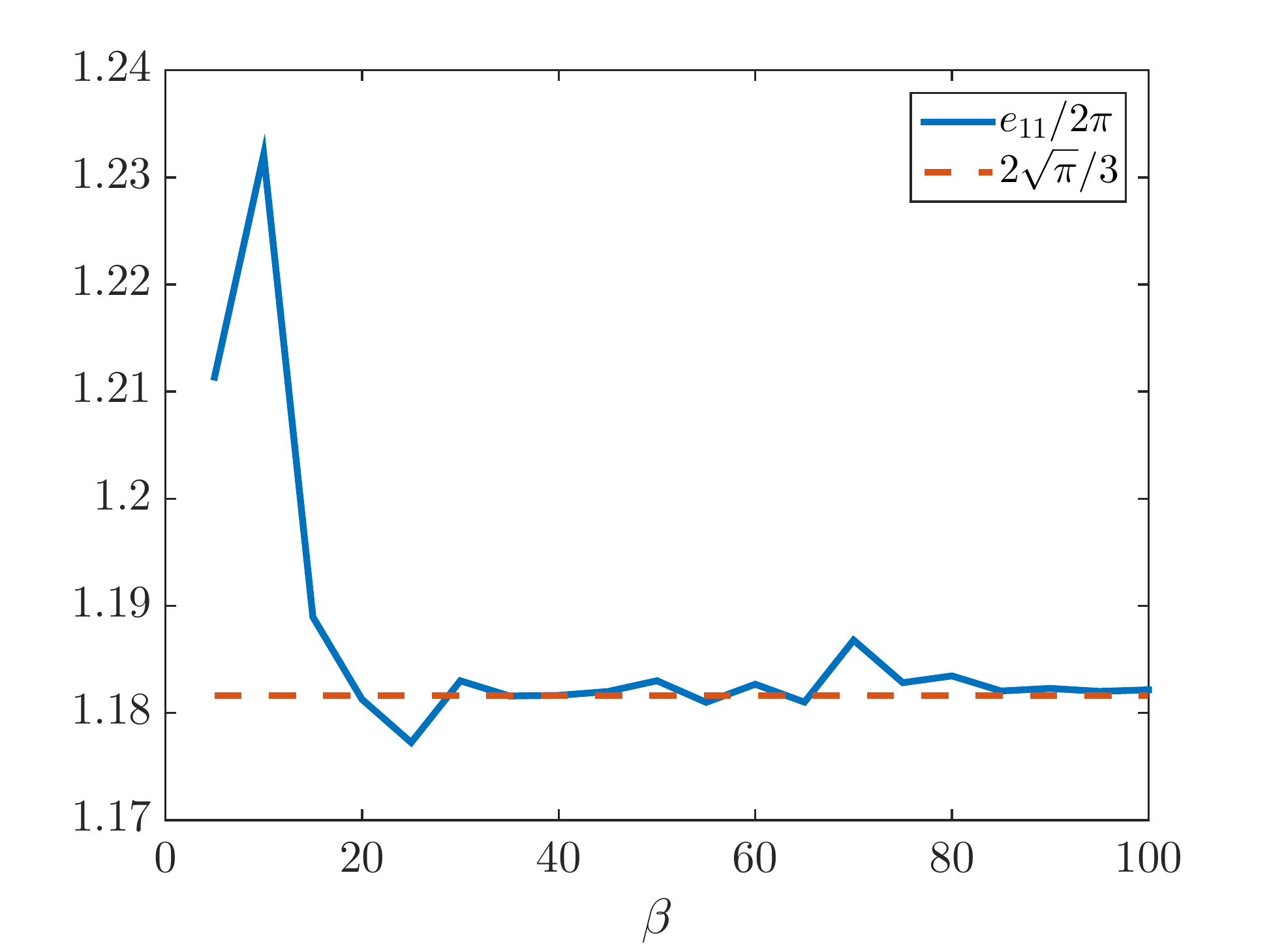}
 \includegraphics[width=5.5cm]{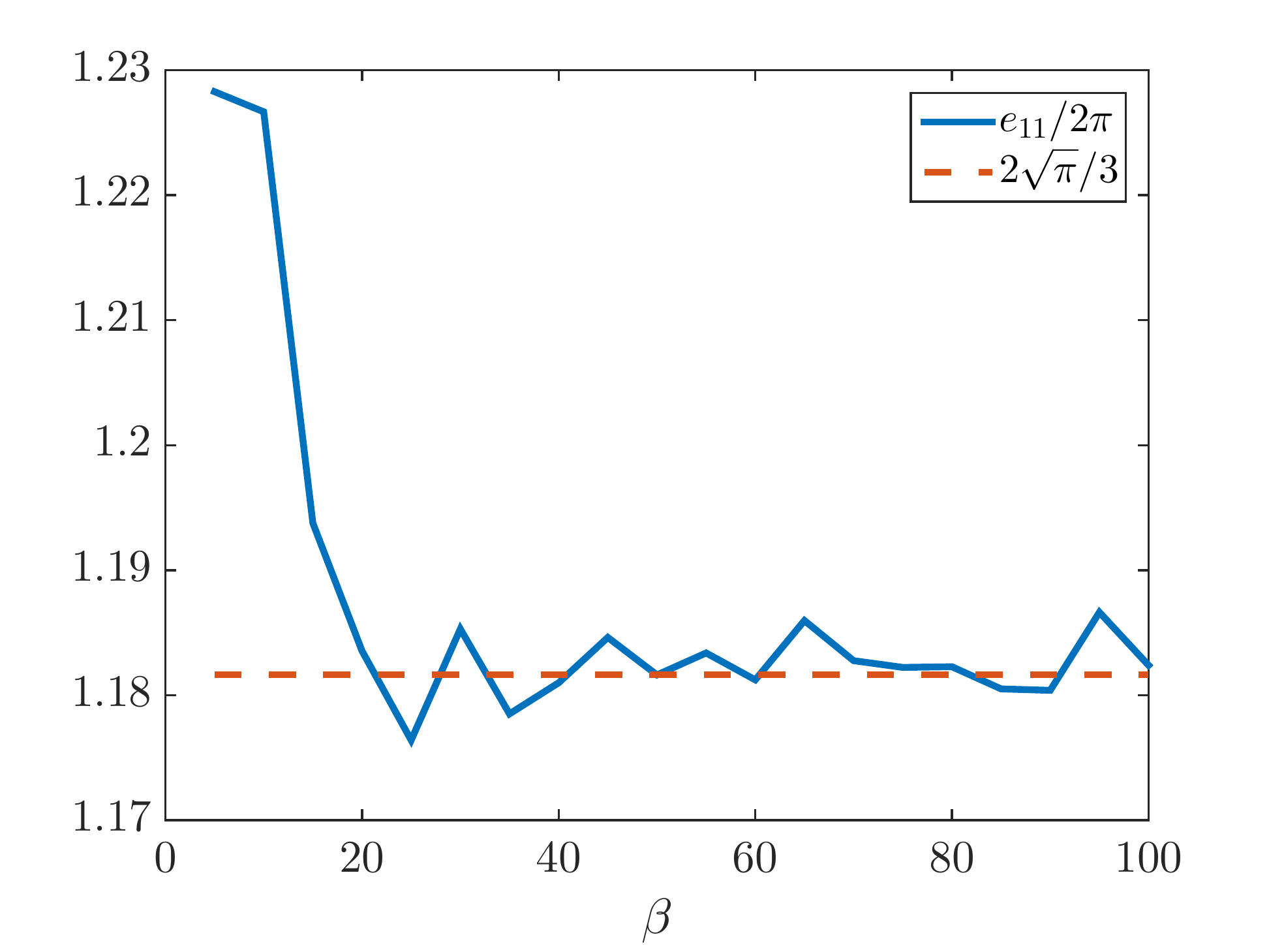}
\caption{Determination of the thermodynamic energy  $e(1,1)$ of the homogeneous mean-field anyon gas at density $\rho = 1$ and total magnetic flux $\beta = 1$. Extrapolated from numerical data with quadratic (above) and quartic (below) traps.}
\label{fig:ener}
 \end{center}
 \end{figure}

For quantitative tests we first extract from the numerical data the value of the unknown parameter $e(1,1)$ using~\eqref{eq:e11 estimate general}. In Figure~\ref{fig:ener} we plot the value of $e(1,1)$ so obtained, as a function of $\beta$ for two types of traps, quadratic and quartic. The convergence for large $\beta$ to the same value for the two traps is a clear sign of agreement between theory and numerics. 

Note that we find $e(1,1) \approx 2\pi \times 1.18 \approx 2 \pi \times (2\sqrt{\pi}/3) $, which is \emph{different} from a frequently used~\cite{ChiSen-92,IenLec-92,Trugenberger-92b,Trugenberger-92,WenZee-90,Westerberg-93} first guess  which is as follows. If one imagines the homogeneous gas as bosons in the lowest Landau level of the approximately constant magnetic field of intensity $2 \pi \beta \rho$, the energy per unit area is $e (\beta,\rho) = 2 \pi \beta \rho ^2$ per particle, hence $e(1,1) = 2\pi$. 

\begin{figure}
\begin{center}
 \includegraphics[width=5.5cm]{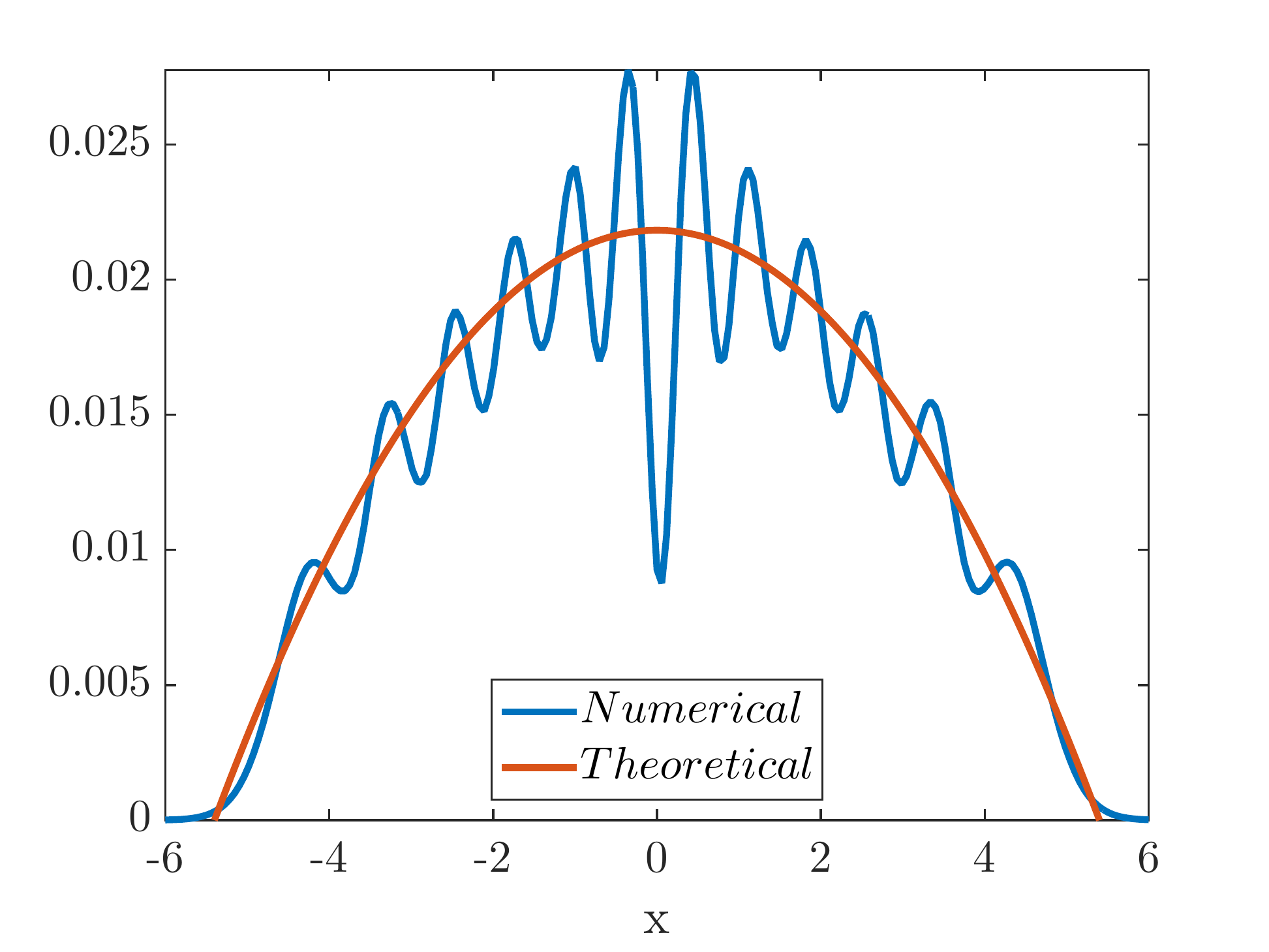}
 \includegraphics[width=5.5cm]{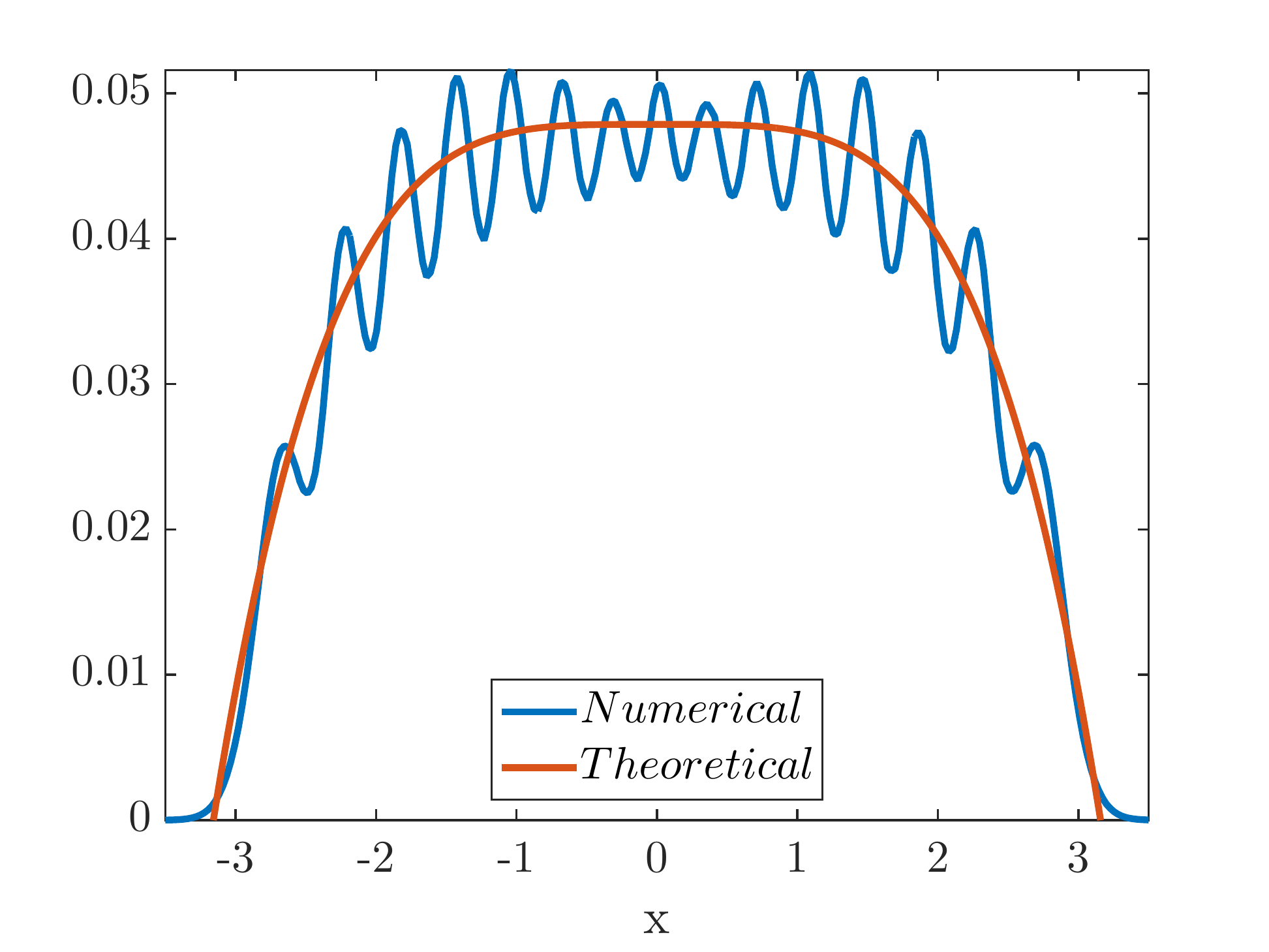}
\caption{Theoretical (red) and numerical (blue) rotationally-averaged density profiles for the mean-field anyon gas. Top panel: quadratic trap $V(\bx) = |\bx|^2,$ magnetic flux $\beta = 90$. Bottom panel: quartic trap $V(\bx) = |\bx|^4$, magnetic flux $\beta = 140$.}
\label{fig:dens prof}
 \end{center}
 \end{figure}

A further benchmark is given by a comparison of the theoretical and numerical density profiles. Now that $e(1,1)$ has been extracted from the numerically computed energies, we can use its value to compute the TF density profile~\eqref{eq:TF profile}. For a reliable comparison with the numerical densities, one must first average out the fast oscillations of the latter, due to the vortex lattice. Comparisons of the theoretical TF profile and rotationally-averaged numerical profile are shown in Figure~\ref{fig:dens prof}. We find an excellent agreement, in particular as regards the essential size of the cloud (TF radius).

\medskip

\begin{figure}[]
\begin{center}
 \includegraphics[width=5.5cm]{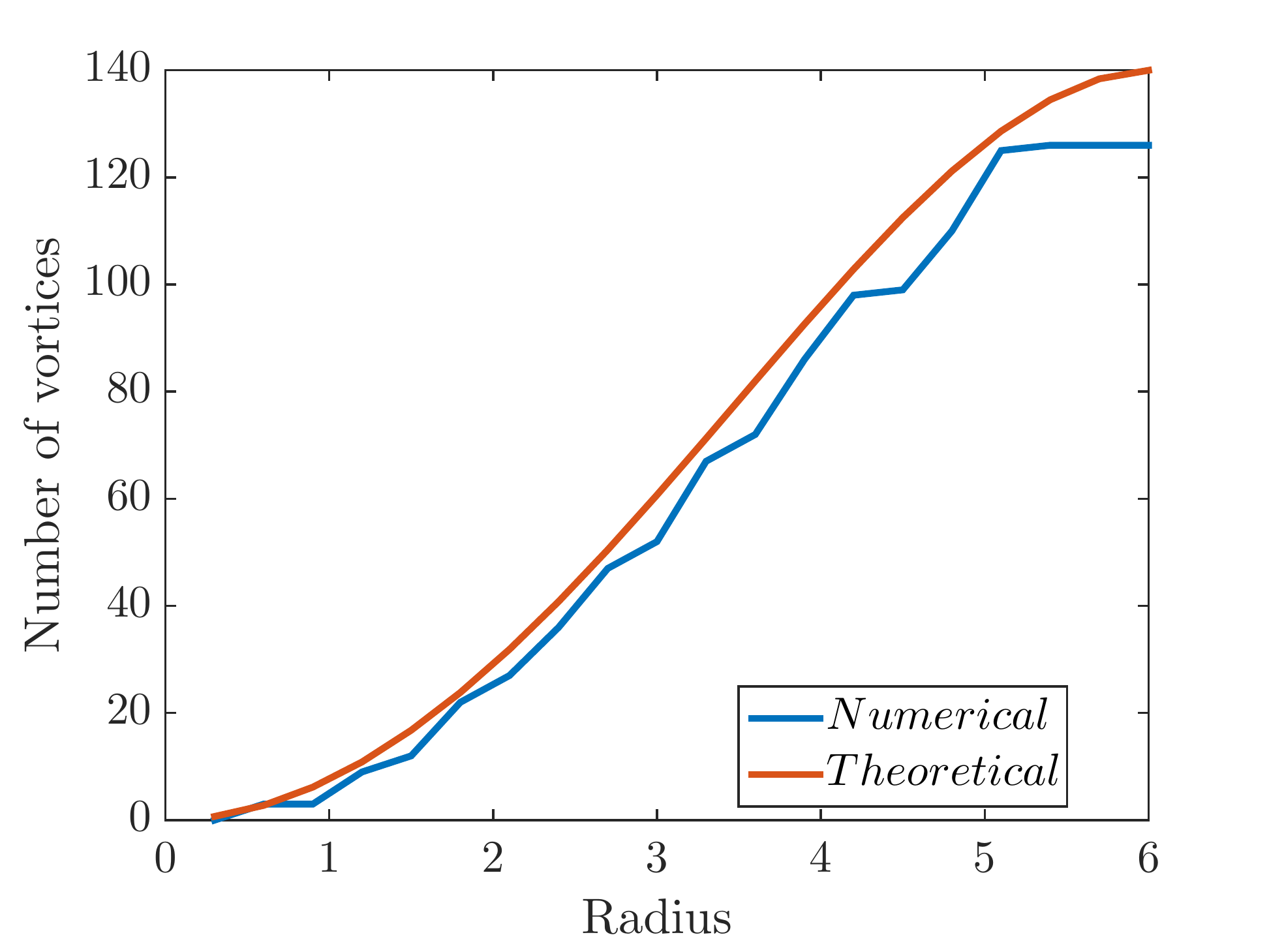}\\
 \includegraphics[width=5.5cm]{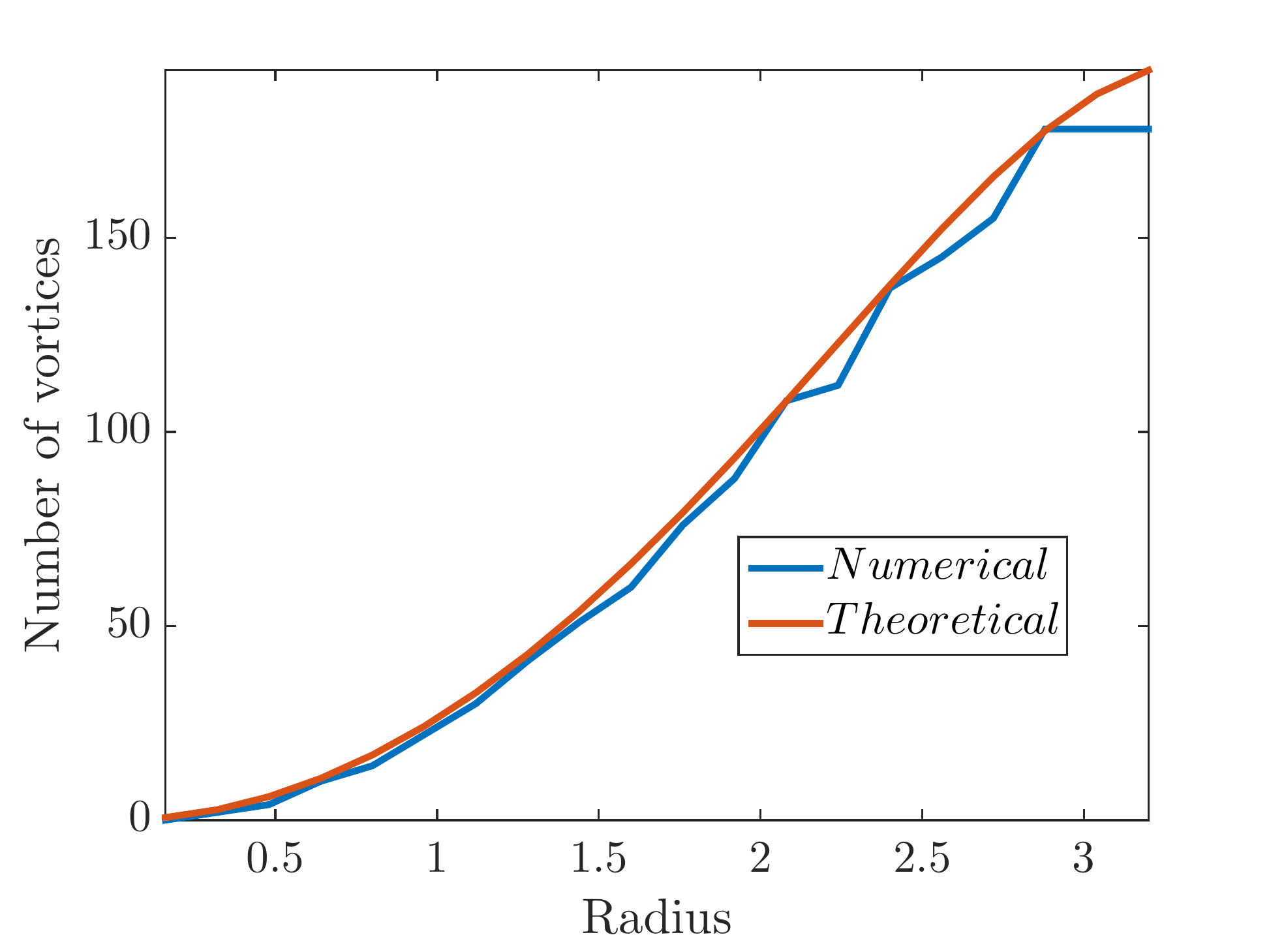}\\
\caption{Theoretical (red) versus numerical (blue) vortex densities in the mean-field anyon gas. Top panel: quadratic $V(\bx) = |\bx|^2$, magnetic flux $\beta = 140$. Bottom panel: quartic trap $V(\bx) = |\bx|^4$, magnetic flux $\beta = 195$.}
\label{fig:vort dens}
 \end{center}
 \end{figure}

\begin{figure}[h]
\begin{center}
 \begin{minipage}{4.25cm}
\includegraphics[width=4.25cm]{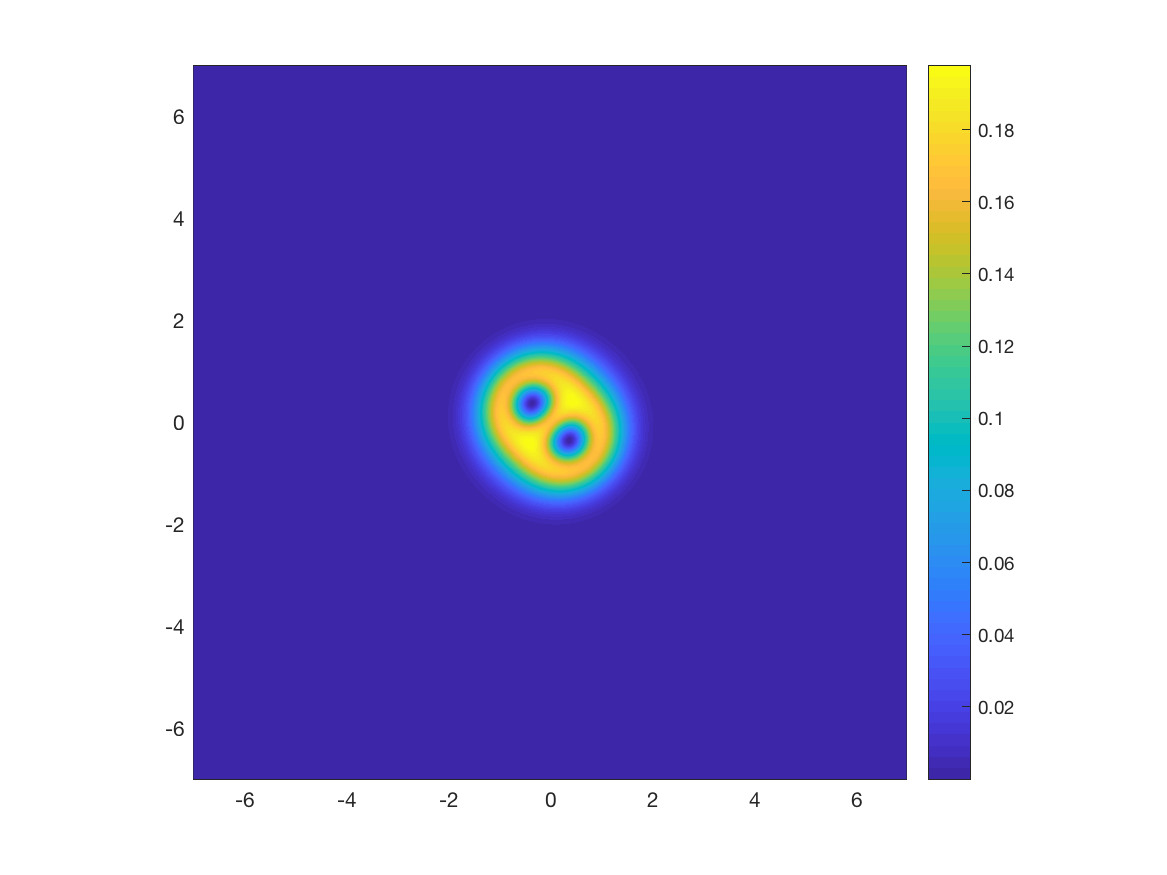}\\
Density, $\beta = 5$
\end{minipage}
\begin{minipage}{4.25cm}
\includegraphics[width=4.25cm]{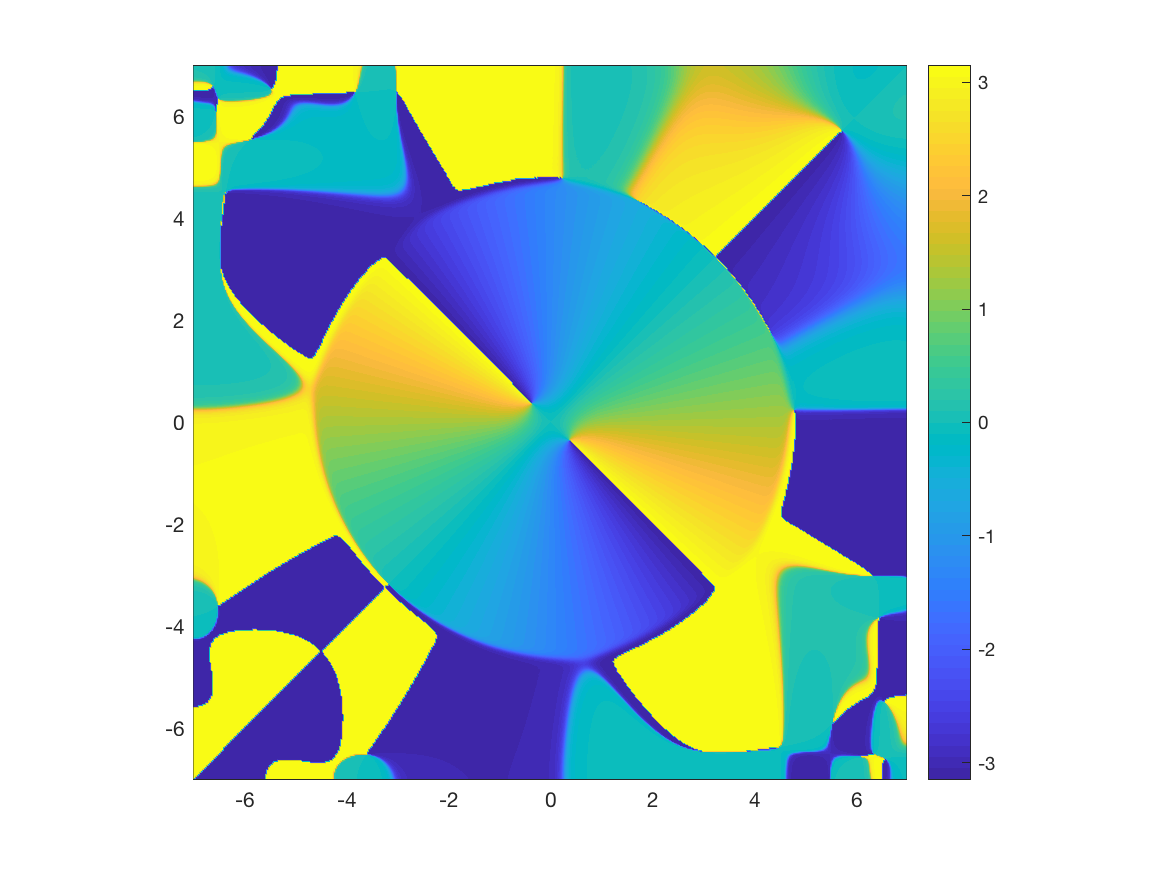}\\
Phase, $\beta = 25$
\end{minipage}
\begin{minipage}{4.25cm}
\includegraphics[width=4.25cm]{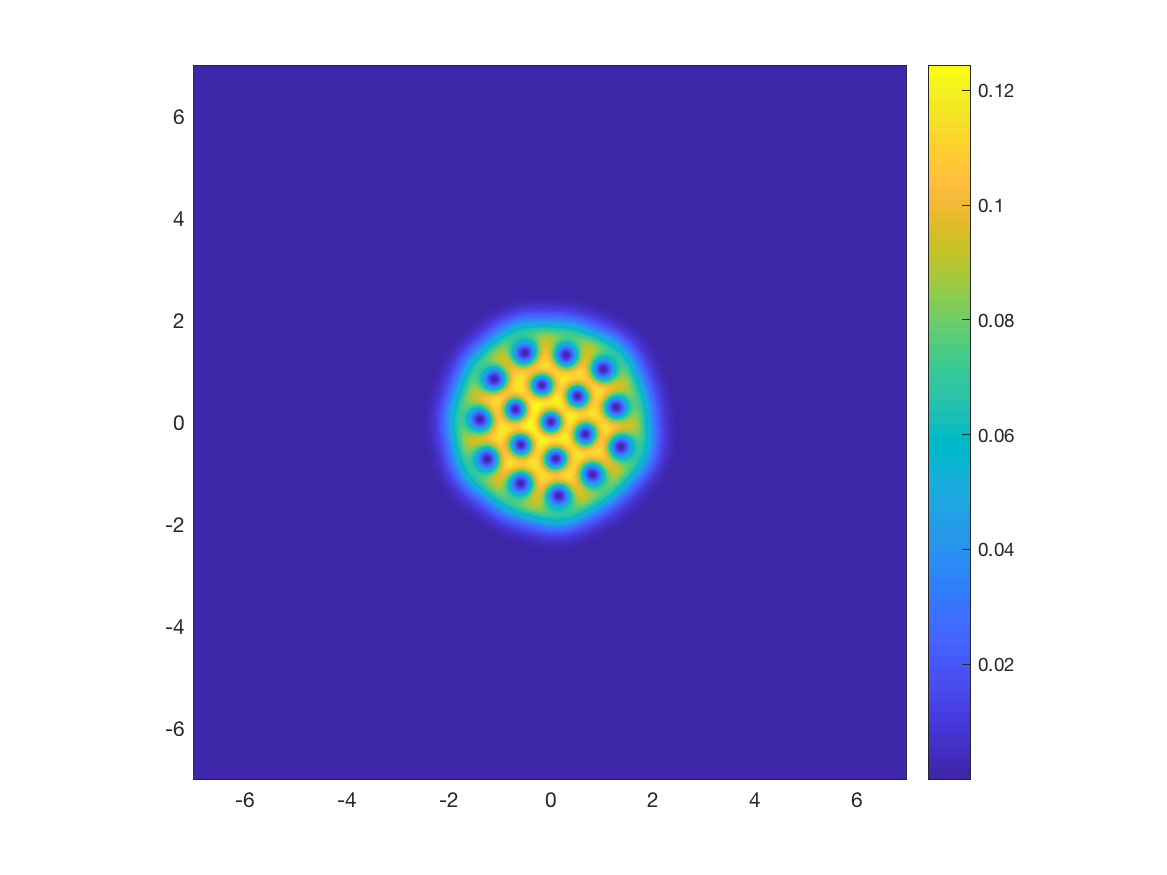}\\
Density, $\beta = 25$
\end{minipage}
\begin{minipage}{4.25cm}
\includegraphics[width=4.25cm]{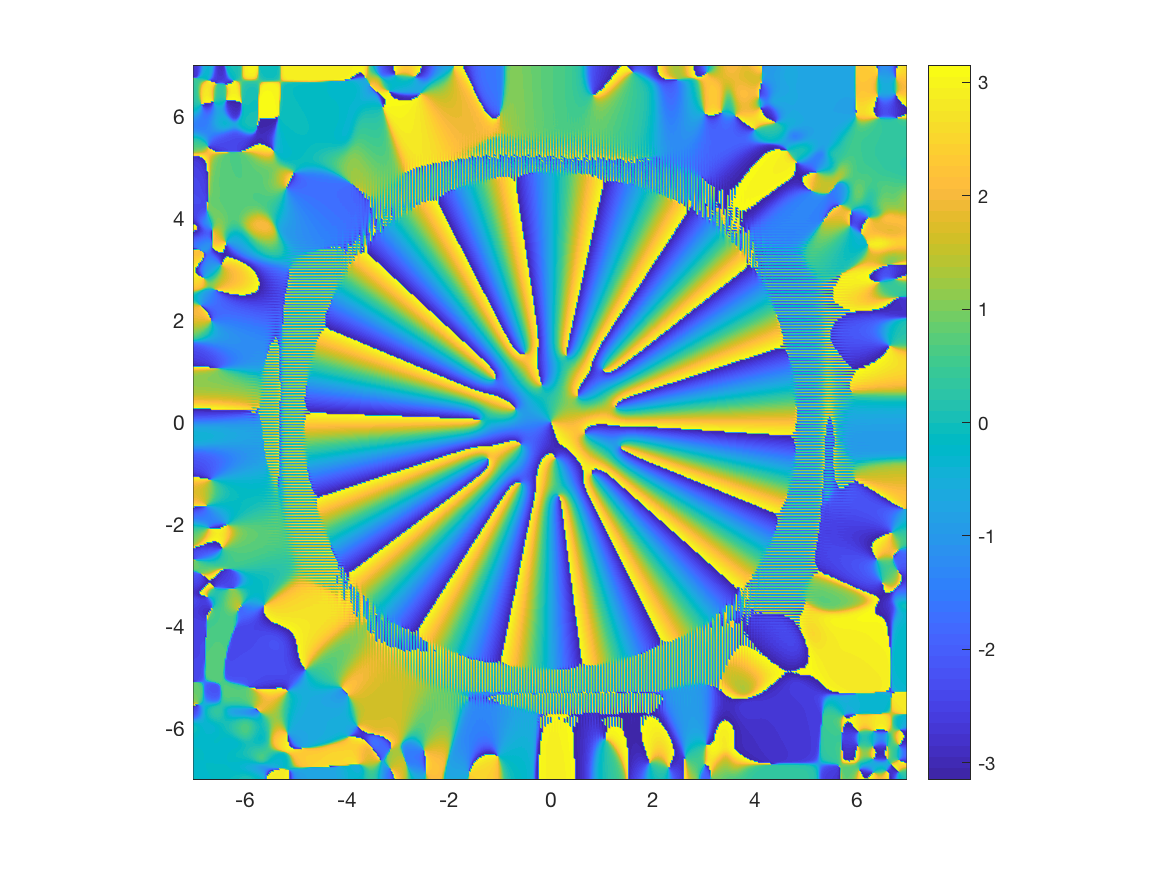}\\
Phase, $\beta = 25$
\end{minipage}
\caption{Density (left) and phase (right) of mean-field approximations $\uaf$ for the almost-bosonic anyon gas  in a quartic trap $V(\bx) = |\bx| ^4$. Top panel: magnetic flux $\beta=5$. Bottom panel, magnetic flux $\beta = 25$.}
\label{fig:phase}
 \end{center}
 \end{figure}

\noindent\textbf{Numerical results: vorticity.} The vortex lattice in Figures~\ref{fig:dens quad} and \ref{fig:dens quart} is clearly visible, as expected. For large values of $\beta$, its inhomogeneity becomes apparent, see in particular the plot for $\beta = 140$ of Figure~\ref{fig:dens quad} where vortices are much more tightly packed at the center of the cloud than at the boundary. For a more quantitative test, we can compare the numerical data to the expected finding~\eqref{eq:vorticity}. More precisely we count the number $N^{\rm num}_v(r)$ of vortices (spotted as zeros in the density) present on the numerical figures within a disk of radius $r$ and compare this to the expected number
$$ N^{\rm theo}_v (r) = 4\pi ^2 \beta \int_{0}^r \rhoTF (s) s ds.$$
Plots of the theoretical and numerical quantities are provided in Figure~\ref{fig:vort dens}. Note that the expected single quantization of vortices may readily be checked by plotting the phase of the wave function,~see Figure~\ref{fig:phase}.

 \medskip

\noindent\textbf{Conclusions.} We have studied a mean-field approximation of the ground state of the many-anyons gas, valid for ``almost bosonic anyons''. A local density approximation suggests a Thomas-Fermi type density profile, whose emergence is due to the screening of long-range magnetic interactions. This and further theoretical considerations suggest the nucleation of vortex patterns in the gas. Our numerical simulations confirm this picture and are found to be in very good agreement with the local density approximation. In particular, the vortex density is directly linked to the matter density, and thus inhomogeneous. 

\medskip

\noindent\textbf{Acknowledgments.} We acknowledge funding from the ERC under the Horizon 2020 Research and Innovation Programme (grant CORFRONMAT No 758620, NR), from the G\"oran Gustafsson Foundation (grant No 1804, DL) and from the Swedish Research Council (grant No 2013-4734, DL).

%
%
%

\end{document}